# From Domain Understanding to Design Readiness: a playbook for GenAI-supported learning in Software Engineering

Rafal Wlodarski
Electrical & Computer Engineering
Carnegie Mellon University Silicon Valley
Mountain View, CA, USA
rafal.wlodarski@sv.cmu.edu

## ABSTRACT

Software engineering courses often require rapid upskilling in supporting knowledge areas such as domain understanding and modeling methods. We report an experience from a two-week milestone in a master's course where 29 students used a customized ChatGPT (GPT-3.5) tutor grounded in a curated course knowledge base to learn cryptocurrency-finance basics and Domain-Driven Design (DDD). We logged all interactions and evaluated a 34.5% random sample of prompt–answer pairs (60/≈174) with a five-dimension rubric (accuracy, relevance, pedagogical value, cognitive load, supportiveness), and we collected pre/post self-efficacy. Responses were consistently accurate and relevant in this setting: accuracy averaged 98.9% with no factual errors and only 2/60 minor inaccuracies, and relevance averaged 92.2%. Pedagogical value was high (89.4%) with generally appropriate cognitive load (82.78%), but supportiveness was low (37.78%). Students reported large pre–post self-efficacy gains for genAI-assisted domain learning and DDD application. From these observations we distill seventeen concrete teaching practices spanning prompt/configuration and course/workflow design (e.g., setting expected granularity, constraining verbosity, curating guardrail examples, adding small credit with a simple quality rubric). Within this single-course context, results suggest that genAI-supported learning can complement instruction in domain understanding and modeling tasks, while leaving room to improve tone and follow-up structure.

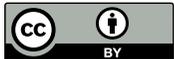

## 1 Introduction

Software engineering education requires both modeling methods and application-domain understanding [1, 2]. Curriculum guides and bodies of knowledge [3, 4] recommend covering a broad range of topics – ranging from SE models and methods to Domain Engineering – emphasizing the breadth of effective SE. Yet delivering this breadth at scale is hard through traditional lecture-based methods as interaction, frequent feedback, and individualized support become impractical [5, 6].

The rise of intelligent tutoring systems, and more recently LLM-based tutoring tools promise scalable, on-demand explanations and examples. Tutoring that relies on conversational exchanges has a well-documented positive impact on student achievement [7, 8] and it is widely available to students with generative AI (genAI). Software Engineering is one domain where genAI tutors proliferated in recent years [9, 10, 11, 12]. Most deployments focus on programming support; far fewer examine supporting skills and knowledge areas—like modeling and domain knowledge—that contribute to classroom readiness.

Early reports cite timely feedback and scalability [13], convenience and reduced discomfort when discussing student knowledge gaps [12], and potential to enhance the personalization, interactivity, and adaptability of learning experiences [14]. However, the instructional quality of genAI tutoring remains under-evaluated, with concerns about reliability and pedagogical fit [15]. Therefore, researchers and educators [16, 17, 18] call for deeper evaluation of genAI-based tutoring.

Motivated by this gap in genAI research for educational purposes, this paper reports on a classroom deployment where 29 master's students in a SE course used a customized ChatGPT tutor to upskill in finance domain knowledge and Domain-Driven Design (DDD) for a team project. We logged all student–genAI interactions and evaluated a ~34.5% random sample (60 prescribed prompt–answer pairs), evenly allocated across sessions, with a multi-dimensional rubric to assess tutoring answer quality. In the paper we aim to address three Research Questions (RQ):

- **RQ1:** To what extent did genAI-supported learning provide high-quality support for students' design-milestone readiness?

- **RQ2**: How do pedagogical value, cognitive load, and supportiveness characterize the quality of genAI-supported learning in a course setting?
- **RQ3**: Which teaching practices enhance the effectiveness of genAI-supported learning in a course setting?

Our contributions are:

- An answer-quality evaluation of genAI tutoring using a practical rubric (accuracy, relevance, pedagogical value, cognitive load, supportiveness) on real classroom data.
- Actionable teaching practices that improved tutoring effectiveness in this setting (prompt design and course-level practices).
- A replicable workflow for collecting, sampling, and analyzing genAI tutoring interactions in SE courses.

Finally, to interpret both answer quality and learner–tutor interaction, we draw on three lenses from learning theory—Cognitive Load Theory (incl. redundancy) [24] to reason about verbosity, jargon, and example count; ICAP [26] to motivate structured follow-ups that move engagement beyond 'active'; and self-efficacy as a complementary learner outcome. These lenses inform our rubric (cognitive load, pedagogical value, supportiveness) and the practices we propose.

The remainder of the paper is structured as follows: first we discuss the study context and how genAI was integrated in the syllabus. We then elaborate on methodological aspects, results obtained and discuss our findings. We conclude by outlining threats to validity and future work.

## 2 Study and context

### 2.1 Course

The classroom deployment took place in a 12-unit master's course on functional programming during Fall 2024. It is a core subject of the Software Engineering program at the university, located in California, United States. The class included in-person sessions twice per week, each lasting 1 hour and 50 minutes. High-level objectives of the course include introducing students to *functional programming* in a hands-on manner and demonstrating how it can be used in industry-relevant settings.

As part of the course, students worked in teams of 3-4 on a project component, which constitutes the context of this study. The project aimed at delivering a realistic automated trading system called "ArbitrageGainer". The project focused on cryptocurrency arbitrage, requiring students to understand market mechanisms to model requirements.

### 2.2 Study environment

To foster effective delivery, the project was divided into four milestones, each lasting two weeks. The first milestone focused on domain understanding and modeling, requiring students to translate functional requirements into a Domain-Driven Design (DDD) model. Subsequent milestones focused on implementation and are not part of this study.

To create a model of the *ArbitrageGainer* app, the *Domain-Driven Design* (DDD) approach was employed. DDD is an industry-relevant software modeling technique that emphasizes aligning the structure and language of software models with the core business domain to improve communication and collaboration between technical and domain experts. The method was introduced to students step-by-step through mini lectures, roughly 30 min long each, where the instructor explained relevant concepts based on a slide deck. During each session, the theory was followed by in-class hands-on activities, where students applied the newly learned concepts to create the design of the project (using a Miro board)

### 2.3 Participants & demographics

29 students completed the course and voluntarily participated in this study. A major share of them had no to little prior knowledge in the areas that were targeted with the described genAI upskilling workflow (Fig.1 ). A handful of students that exhibited more significant finance experience, stated they either had their own investments and trading activity on the stock market (5), worked in the fintech sector (2) or are a finance/economics minor (1). Significant DDD acumen was a little less prevailing among students and stemmed from prior coursework (3) or industry experience (2).  With regards to the undergraduate studies completed, more than half of the students completed a Computer Science degree (15), followed by Electrical and Computer Engineering (4), and Software Engineering (3). Graduates from other engineering disciplines and finance/math related majors were also present (7).

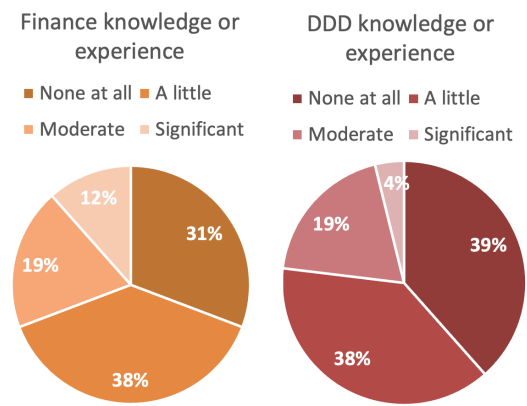

Figure 1: Students' self-reported background

## 3 Integrating Generative AI in the syllabus

This section describes the intervention as students experienced it (schedule, tasks, prompts, and tutor setup).

With students arriving from diverse backgrounds—and limited instructor bandwidth during 1h50m in-person sessions—guiding each team through domain knowledge gaps, functional-requirements misunderstandings, and Domain-Driven Design method became untenable. Although class size wasn't huge in the previous course instance (34), circulating among teams left many questions unanswered until after class, when the instructor would spend 10–30 extra minutes with lingering groups. In response, we designed a genAI-enabled upskilling workflow to close that support gap and strengthen students' mastery of supporting SE knowledge areas. Although this gap could, in principle, be mitigated by teaching assistants, such support depends heavily on their competencies and reliability. In our case, despite being explicitly tasked with in-class assistance, the TA did not attend, leaving the professor to manage the course block alone.

### 3.1 Course design

To effectively introduce the genAI-enabled upskilling workflow, some tweaks to the course were necessary. Firstly, we wanted to set a proper frame for use of genAI as students might view the genAI activities as extra work rather than a helpful tool, affecting their attitude and engagement. Therefore, an explicit Learning Objective on using genAI was added (LO-genAI): "*Use generative Artificial Intelligence (genAI) to obtain domain knowledge and properly apply domain-driven design on a collaborative project*". Furthermore, the how and why of using genAI in the course was mentioned during the first lecture and repeated before the intervention took place. Finally, all the genAI-based activities were made mandatory and were rewarded with course credits. Based on prior course iterations, optional preparatory work often saw uneven uptake, with stronger students engaging more and others skipping it entirely; making genAI activities mandatory and credit-bearing was intended to ensure consistent participation across the cohort.

**GenAI literacy among students.** At the time, genAI use on campus was expanding rapidly but in an unstructured way, with students entering the course with widely varying experience. To establish a common baseline, the university required all students to complete a genAI literacy module during the first week of class. This asynchronous module, worth 1% of the course grade, took 90–120 minutes and included a knowledge check, four short learning modules, and a knowledge review. The learning objectives focused on understanding how genAI tools work, recognizing ethical and contextual considerations, and applying responsible use strategies in educational tasks.

### 3.2 GenAI-enabled upskilling workflow

We identified two knowledge areas that repeatedly limited workshops progress: (i) cryptocurrency-arbitrage domain concepts and (ii) DDD concepts required for modeling. To improve readiness for in-class workshops, we introduced mandatory pre-class genAI upskilling tasks aligned to the next workshop step, followed by a short individual knowledge check/reflection report. For example, in DDD, one modelling step involves identifying external systems (outside the solution domain and development scope). This proved unexpectedly difficult as students failed to extract systems such as cryptocurrency exchanges and market data API from the functional requirements, even though proprietary names and usage were explicitly stated. To address this, we added a prompt asking about typical subsystems and components in an automated trading system, paired with a knowledge check to ensure students considered external systems explicitly. While some individual reports still contained errors, teams ultimately identified all external systems correctly, and no mistakes remained in the final deliverables.

Following every upskilling activity (executing verbatim a predefined set of prompts), a student was required to submit a report. It consisted of knowledge check questions of varying difficulty and effort required to answer them which validated student's understanding of the terms and concepts covered. Additionally, for research purposes, reports included some reflection elements requiring students to analyze their interaction with genAI and share a specific insight (each report inquired about a different aspect). The reports were graded manually by the instructor and were qualified as pass if (i) all knowledge-check items were answered correctly with brief reasoning and (ii) the transcript was attached. These design decisions were driven by timeliness of feedback necessary (report was to be graded before the upcoming class to validate student's readiness) and the fact that the upskilling activities (3) were condensed within 1,5 week period. ~93% of students provided their reports within the expected timeframe and passed all three reports on the first submission, indicating baseline readiness before in-class work.

**Table 1: Two-week genAI upskilling schedule (student-facing).** Pre-class tasks were completed individually; in-class workshops were team-based. Prompts are provided in replication package.

| Week | Session & workload | Objective | Student tasks/activities | Artefacts produced |
|---|---|---|---|---|
| 1 | Pre-class #1 *(individual, 30-45min - estimate)* | Finance domain knowledge bootstrapping | Read project spec; run prompt 1.1; run prompt 1.2; optional follow-ups; export transcript; complete Knowledge Check (KC) #1 | Transcript of 2 prescribed prompt–answer pairs (+ optional follow-ups); KC#1 report (3 open-ended questions) |
| 1 | In-class #1 *(lecture + team workshop, 1h50)* | Project and DDD modeling kick off | System overview; DDD introduction lecture; event storming (team); | Miro board with events (team) |
| 1 | Pre-class #2 *(individual, 20–30 min - estimate)* | Familiarity with DDD concepts | Run prompt 2.1; run prompt 2.2; optional follow-ups; export transcript; complete KC#2 | Transcript of 2 prescribed prompt–answer pairs (+ optional follow-ups); KC#2 report (2 open-ended questions) |
| 1 | In-class #2 *(team workshop, 1h50)* | Key DDD activity completed | Event storming continued: derive timelines/commands/external systems (team) | Updated Miro board with ordered events, commands, external systems |
| 2 | Pre-class #3 *(individual, 30–45 min - estimate)* | DDD terms application to project deliverables | Run prompt 3.1; run prompt 3.2; optional follow-ups; export transcript; complete KC#3 | Transcript of 2 prescribed prompt–answer pairs (+ optional follow-ups); KC#3 report (1 open-ended question) |
| 2 | In-class #3 *(lecture + team workshop, 1h50)* | Milestone deliverables kick off | Modeling workflows and documenting a business domain lecture; workflows definition (team); bounded contexts definition (team) | Updated Miro board with bounded contexts; draft of workflows pseudocode documentation (final project deliverable) |

The upskilling workflow was executed throughout the milestone as outlined in Table 1.

### 3.3 Prompt design & genAI set up

Key to the intervention was defining and administering genAI prompts that students were to execute at specific points in time to support their learning. The prompt set was established to support the course's learning objectives: (LO-DDD) "*Design a Domain-Driven model for a real-world system*", and LO-genAI (see 3.1). Each session's two prescribed prompts were mapped to the next in-class workshop step (e.g., domain terms clarification catered to correctness of event storming). Prompts were derived from recurring misconceptions observed in prior course offerings (milestone artifacts, in person and course forum questions raised). Other design considerations included time effort required to execute a prompt and analyze the output as well as appropriateness of the material to be conveyed in a conversational format.

GenAI tutor itself was tailored to both the course and the type of tasks it was meant to assist the students with. Customization was two-fold:
1) Using a tailored, detailed prompt to frame the tutor's profile, problem and the task.
2) Using grounding via uploaded course materials (slides, project specification, exemplar solution) provided through the ChatGPT knowledge configuration.

The instructions for a genAI model included the following elements as per good practices concerning prompt engineering [1]: a persona to provide context, pointers to course materials, description of tasks to be performed, a list of constraints, "dos and don'ts" as well as desirable traits of the answer's format. The prompt can be found in replication package. The knowledge base of the customized genAI included a set of relevant lecture slides (pdf), project specification (functional requirements) and an exemplary student solution (DDD model).

The model was built using GPT-3.5 Turbo, with default parameters (not user-configurable in the ChatGPT interface). The latest model version available at the time was selected for availability and ease of integration; no comparison with alternative models was conducted.

Before the semester, the instructor iteratively prototyped the tutor configuration prompts using scripted runs representing expected student interactions. Prompt wording and/or knowledge-base coverage were revised until responses consistently (across 5 trials) matched an acceptable answer pattern: correct course-aligned terminology, sufficient depth for the upcoming workshop task, and minimal boilerplate. The model was finalized in its 4th version and prompt set was frozen at the start of Session 1 to ensure consistency across students, at the expense of interaction diversity. Students accessed a shared tutor instance without authentication.

## 4 Methodology

This section describes the data collected from the deployment and how we sampled and scored it.

### 4.1 Data sources

Extensive data was collected as part of the deployment at different points in time of the semester.

*4.1.1 Pre-intervention* During the first week, a demographics survey was distributed to get insight into the profiles of students participating in the study. It consisted of standard questions concerning their bachelor's degree completed as well as competencies in course relevant areas, namely functional programming, finance, and Domain-Driven Design. We asked about their knowledge and experience in each area on a 4-level scale (none at all; a little; moderate; a lot) and for each area, a student was asked to describe how the knowledge/experience was gained (prior courses, self-taught, work experience).

We were also interested in students' beliefs and attitudes towards genAI. To establish a baseline, we included a set of 5 self-efficacy statements in the demographics survey. We defined two primary self-efficacy outcomes aligned with genAI use (SE-1 domain knowledge with genAI; SE-2 applying DDD with genAI). A third item on general DDD modeling (SE-3) was treated as exploratory, as gains could stem from hands-on project work independent of AI.

*4.1.2 Intervention* Following every upskilling activity, a student submitted a reflection report and extraction of the conversation with genAI to the course's LMS (Canvas). Students were instructed to execute prescribed prompts to standardize evaluation; optional follow-ups were allowed but excluded from analysis.

*4.1.3 Post-intervention* The last week of classes, during a live session, an end-of-term survey was distributed to collect feedback on the use of genAI in the course as well as probe students' self-efficacy.

### 4.2 Sampling

The corpus comprised 174 prompt–answer pairs: three upskilling sessions × 29 students × two prescribed prompts per session. We define a conversation unit as the contiguous pair of prescribed prompts (two prompt–answer pairs) submitted by a student within a given session, yielding 87 conversation units in total (29 per session). To balance feasibility and coverage, we performed stratified random sampling by session. Using an external, time-seeded randomization tool (random.org), we selected 10 conversation units per session (30 units total, without replacement), which produced 60 prompt–answer pairs (34.5% of all pairs). This design (i) preserves within-session context by keeping the two prescribed prompts together, and (ii) ensures equal representation of each upskilling session, reducing session-level confounding.

Rubric scoring was applied to each prompt–answer pair within the sampled conversations; aggregation is reported at the item and session levels. Because the prescribed prompts were session-specific (Session 1: finance, Session 2: DDD basics; Sessions 3: advanced DDD), stratifying by session guarantees thematic coverage by construction. As a sensitivity check, we recomputed rubric means by session and with any single session omitted; the qualitative conclusions (e.g., very high accuracy/relevance, lower supportiveness) were unchanged.

### 4.3 Evaluation rubrics and scoring procedure

In order to probe usefulness of genAI as an upskilling tool for Software Engineers, we targeted a handful of assessment criteria based on the current state of research as well as educational needs. Firstly, it is a well-known fact that LLMs are prone to hallucinations [27, 28] therefore we tracked "Accuracy of information" fed to students. Secondly, we verified that only "relevant" passages were provided, devoid of off-topic rants. Moving to the educational aspects, as defined by Paul Denny et al. [21] genAI Teaching Assistants in Programming Education should be characterized by scaffolding (not providing solutions) and appropriateness (avoiding jargon, using plain language). In our study we mapped those qualities to "Pedagogical value" (defined as well-structured responses containing thematical depth and structuring examples) as well as "Cognitive load" (which encompasses clarity, conciseness and accessibility). Finally, we were interested to see if genAI feedback is consistently supportive and validating to students – this aspect was captured by "Support and encouragement" dimension. "Supportiveness" is an author-defined proxy operationalized as brief praise/validation plus an invitation to explore next steps; it is not a validated affect scale. The resulting rubric consisted of 5 dimensions deemed important to answer our RQs and can be found in Table 2. Scores were produced by the course instructor—arguably the most qualified domain judge in this context.

**Rater agreement check.** Two trained TAs who took the course previously independently rated the same subset of 15 transcripts (5 per session) after a 90-minute rubric calibration on 5 anonymized transcripts. Agreement with instructor ratings on the overlap set (n=150 paired ordinal ratings) was high: weighted Cohen's κ = 0.76 (TA1) and 0.78 (TA2), with 72–74% exact agreement and 99.3% within-one agreement (|Δ|≤1).

**Replication package** We provide a replication package containing the prescribed prompts, tutor configuration, rubric materials, de-identified transcripts/ratings (where permitted), and analysis scripts as supplementary material with this submission.

## 5 Results

### 5.1 Answer quality and perceived effectiveness of genAI-supported learning

To answer RQ1, we assess support through two quality dimensions: "Accuracy of information," capturing factual correctness and absence of conceptual errors, and "Relevance," reflecting how well responses address the student's question with precise, context-aware information. Our dataset was characterized by a near perfect accuracy of the information provided (average accuracy score of 98.9% across probed conversations). None of the data points contained hallucinations or false information. Only 2/60 (3.3%) responses had "minor inaccuracies that do not significantly mislead the student" meaning that they do not impact the learning experience negatively.

Table 2: Rubric for evaluating feedback quality in relation to RQ1 and RQ2

| Criteria | Exemplary (3) | Proficient (2) | Developing (1) | Unsatisfactory (0) |
|---|---|---|---|---|
| **Accuracy of Information** | The response is factually correct, aligns with domain knowledge, and contains no conceptual or technical errors. | Mostly correct but may include minor inaccuracies that do not significantly mislead the student. | Contains noticeable inaccuracies or outdated information that could lead to misunderstandings. | Contains significant errors or misconceptions, making the response misleading or incorrect. |
| **Relevance** | The response directly addresses the student's question with precise, context-aware information. | Mostly relevant but may include minor off-topic details or incorrect assumptions. | Somewhat relevant but misinterprets the student's question or include substantial off-topic content. | Does not address the student's prompt or is completely off-topic. |
| **Pedagogical value (depth & learning contribution)** | The response provides structured reasoning, correct terminology, and relevant examples, ensuring sufficient depth for student understanding. | The response includes useful learning elements/examples but lacks depth in explanation, or omits key details. | The response is superficial, missing important details, examples, or structured reasoning necessary for learning. | The response lacks instructional value, providing little to no meaningful explanation or structure, , or relevant details. |
| **Cognitive load (clarity, conciseness and accessibility)** | The response is well-structured, concise, and easy to understand while avoiding unnecessary complexity. Language is appropriate for the student's level. | Mostly clear and accessible, but may be slightly wordy, redundant, or require some effort to interpret. | Somewhat difficult to understand due to lack of structure, verbosity, ambiguous phrasing or overwhelming in terms of information quantity. | Unclear, disorganized, overly technical or general overload of information making comprehension difficult for the student. |
| **Support and Encouragement** | Feedback is consistently supportive and validating. It recognizes the student's effort, provides genuine encouragement to explore the topic further, and fosters a positive learning mindset. | Feedback includes some encouraging elements but may feel somewhat formulaic or inconsistent. It acknowledges the student's effort but lacks the depth or personalization to fully inspire further inquiry. | Feedback provides minimal acknowledgment or positivity toward the student's effort. It is not explicitly discouraging, but lacks warmth, genuine praise, or motivational language. | Feedback is dismissive, negative, or overtly critical. It fails to validate the student's question or motivation, discouraging continued engagement. |

With regards to the relevance of information provided, genAI tutoring was characterized by very high scores (average of 92.2%). None of the responses scored below 2 (out of 3) on the rubric used, while 82.5% of the responses obtained a maximal score. Taken together, the accuracy (98.9%) and relevance (92.2%) scores demonstrate high answer quality of genAI tutoring in this course context. We treat these dimensions as proxies for tutoring effectiveness—reflecting the quality of support provided—rather than direct evidence of learning outcomes.

Therefore, we investigated the research question further by looking at students' self-reported confidence gains in a post-session survey. We treat self-efficacy as complementary to answer-quality evidence. In a paired pre–post analysis, all three self-efficacy items showed statistically significant gains. Item 1 ("*I can use genAI to obtain domain knowledge*") increased from pre to post, $F(1,22)=24.41$, $p<.001$; the effect was large (partial $\eta^2=0.53$; $d^z\approx1.03$), indicating a practically meaningful boost in confidence for domain learning. Item 2 ("*I can use genAI to properly apply DDD on a collaborative project*") also rose significantly, $F(1,21)=14.81$, $p<.001$, with a large effect (partial $\eta^2=0.41$; $d^z\approx0.82$), suggesting improved confidence in applying DDD with teammates. Item 3 ("*I can design a domain-driven model for real-world business needs*") showed the strongest gain, $F(1,21)=45.82$, $p<.001$; the effect was very large (partial $\eta^2=0.69$; $d^z\approx1.44$), reflecting substantial growth in perceived modeling capability. However, this item does not reference genAI and likely reflects learning-by-doing in the project, we report it as exploratory and do not use it to answer RQ1.

Because tutoring co-occurred with mini-lectures and project practice, self-efficacy gains should be interpreted as associated with genAI-supported learning in this setting, alongside other instructional activities.

## 5.2 Pedagogical quality of genAI tutoring

We addressed RQ2 by scoring each genAI response on pedagogical value, cognitive load, and supportiveness, and we summarize the aggregates and recurring failure modes below.

GenAI responses showed high pedagogical value overall (89.4%). Most answers were deep enough to advance learning—77.5% received the maximum score. One answer (1/60) scored 1/3: when asked to list typical systems/components of an automated arbitrage platform, the reply was overly high-level (labels with little explanation). Given the right-skewed distribution and rarity of such cases, we treat this as an outlier rather than a pattern.

Cognitive load was appropriate on average (82.78%). Only 5% of answers were flagged as overwhelming.

Supportiveness—as operationalized in our rubric—was low (37.78%). When not explicitly prompted with a coaching persona, the tutor tended to be neutral and informational, offering little praise or motivational language.

## 5.3 Teaching practices that enhance genAI-tutoring effectiveness

Below we group the practices by where instructors can act: prompt/config, course design, and workflow/assessment. Each item specifies intent (why) and an actionable step (how). The proposed set of practices was derived either from the authors' experience rolling out the genAI tutoring (E – evidence based) or observed failure modes and strengths of the genAI tutor (R – reflective).

*5.3.1 Prompt engineering & AI configuration.*

- **P1 Set expected granularity (R)**. *Why*: Prevents overly shallow or overly detailed answers that misalign with the task. *How*: Specify the target depth in the system prompt (e.g., "explain in 3–4 sentences with one concrete example") and validate against sample questions during prototyping.
- **P2 Constrain verbosity (R)**. *Why*: Reduces unnecessary text (repetition and long preambles) that increases cognitive load without adding value. *How*: Impose explicit limits (e.g., ≤2 short paragraphs) and instruct the model to omit introductions and summaries unless requested.
- **P3 Cap examples (R)**. *Why*: Too many examples can overwhelm students or dilute key points. *How*: Limit to 1–2 concise examples per response, using a single example for simple concepts and at most two for more complex ones.
- **P4 Curate guardrail examples (R)**. *Why:* Poor or generic examples reduce learning quality. *How:* During prototyping, sample multiple runs, select the best 1–3 exemplars, and embed them in the system prompt as the default examples to use.
- **P5 Block unfamiliar jargon (E)**. *Why*: Unexplained terms increase cognitive load. *How*: Maintain a small "banned" list and require definitions for any term outside the provided materials. Repeat 2 or more times in very long prompts.
- **P6 Add a supportive tone (R)**. *Why:* Default responses lacked encouragement. *How*: Define a tutor persona that acknowledges effort and invites further exploration in each response.
- **P7 Curate the knowledge base (E)**. *Why*: Missing domain coverage led to occasional irrelevant passages. *How*: Include concise materials covering all required domains (core and supporting) in the tutor's knowledge configuration.
- **P8 Signal, don't repeat (R)**. *Why*: Repetition across sections increases cognitive load without adding value. *How:* Use short headings/keywords as signposts (e.g., "Bounded context — definition"), enforce a one-pass rule ("state each key point once; don't rephrase later"), and replace repeats with brief pointers ("*As defined above…*"). No paragraph-length summaries unless explicitly requested.
- **P9 GenAI Nudge (R).** *Why:* Promotes constructive/interactive engagement (ICAP) by eliciting targeted follow-up questions. *How***:** Prompt the model to suggest one specific next step or question (e.g., clarify, apply, or extend the concept) per session. This keeps the dialogue purposeful and supports deeper engagement. [26]

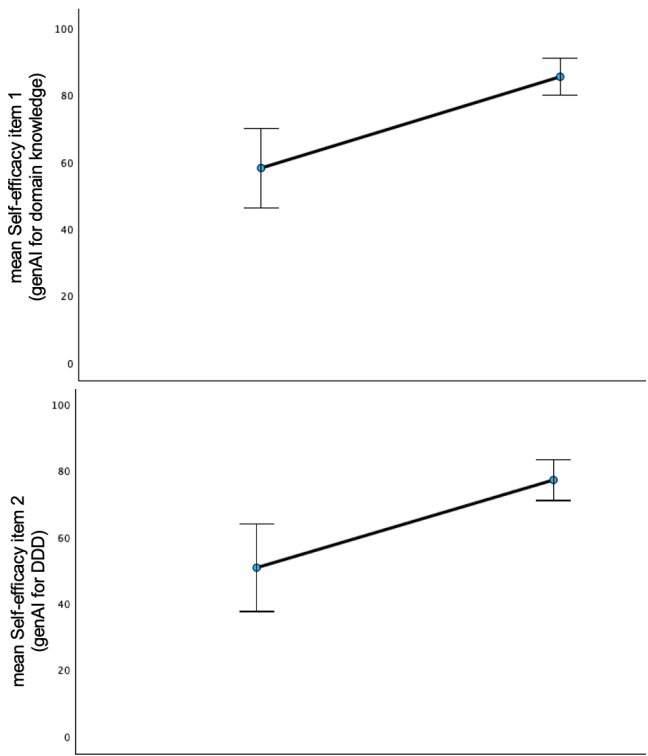

**Figure 2: Self-efficacy before and after the AI-tutoring session.** Points show group means; vertical bars are 95% confidence intervals.

*5.3.2 Course design*

- **P10 Make genAI use an LO (E).** *Why:* Signals relevance and legitimizes effort. *How:* Add a learning outcome explicitly requiring effective use of genAI for analysis or modeling tasks.
- **P11 Front-load a mini workshop (E).** *Why:* Establish a shared baseline of genAI literacy as students lack prompt and evaluation skills. *How:* hands-on 30–45 minutes on prompt patterns, asking follow-ups, and critical evaluation of responses.
- **P12 Credit the work (E).** *Why:* Sustains engagement; our sessions were short but non-trivial. *How:* Allocate 1–5% total course weight proportional to required effort.
- **P13 Centralize guidance (E).** *Why:* Reduces confusion about usage. *How:* Provide a single LMS page with access instructions, prompt examples, and usage expectations.

*5.3.3 Upskilling workflow, assessment & scaling*

- **P14 Require brief notes/logs (E).** *Why:* Enables reflection and usage tracking. *How:* Students export a short conversation log following a template.
- **P15 Use a simple rubric for knowledge checks (R).** *Why:* Distinguishes careful curation from minimal effort. *How:* Three levels per question mapped to quick clicks in the LMS; formative comments optional.
- **P16 Delegate at scale (R).** *Why:* Instructor grading does not scale. *How:* Train TAs on the rubric and calibrate using a small shared sample before grading; batch-grade per question.
- **P17 Vary question difficulty and format (E).** *Why:* Supports different prior knowledge levels. *How:* Design tasks that mix recall (definitions), application (apply to project), and small design exercises (e.g., define one bounded context).

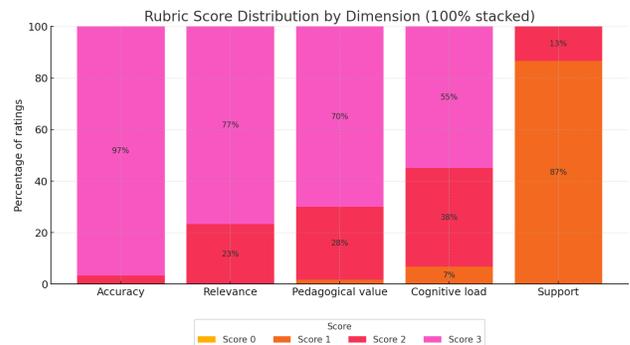

**Figure 3 Percentage distribution of genAI tutoring ratings (0–3) across rubric dimensions.**

## 6 Discussion

### 6.1 RQ1: information quality and self-efficacy

Results reflect answer quality and self-reported confidence; no causal claims about learning are made.

**Accuracy of information.** In our sample of 60 genAI-tutoring responses, accuracy was near-perfect (≈98.9%), with no factual errors and only two responses (1.1%) rated as "minor inaccuracies" that did not mislead learners. This contrasts with common concerns about LLM "hallucinations" [27, 28] and suggests that—in a course-grounded setup—genAI tutoring can deliver reliable explanations. Student reflections aligned with this pattern: (e.g., "*concise and accurate introduction of concepts*"; "*genAI provided accurate information when I struggled*").

**Why it was high and where it slipped.** We hypothesize that constraining the tutor to a curated knowledge base —slides, readings, and specifications— (P7 Curate the knowledge base, §5.3) and prompting it to stay within those materials, likely reduced off-domain drift. Two non-maximal cases appeared late in the upskilling workflow and considered advanced DDD concepts. One echoed a somewhat misleading characterization of a "Domain service" common in secondary web sources (not a pure hallucination), underscoring the importance of curating an exhaustive knowledge base. The second one used an elaborate example with a minor context mismatch (an event name that did not cleanly belong to a stated bounded context). This can be avoided by pre-vetting exemplars (P4 Curate guardrail examples, §5.3) for prescribed prompts so illustrative content remains precise.

Note that the accuracy scale showed ceiling effects and ratings reflect a single expert rater; results should be read as precise for this context, not universal.

**Relevance of information.** Relevance was very high (92.2%), supporting the view that, in a course-grounded setup, the tutor often stayed on-task. The primary enabler was a course-grounded knowledge base (P7), as echoed by students (e.g., "*knowing the project description made answers more concise.*").

Where relevance dipped, three patterns dominated. Gaps in the KB (finance concepts were essential to the project but not fully represented beyond the specification) lead to occasional off-scope passages and usage of terms that were outside of the functional scope of the team project. This can be easily addressed by expanding the KB to cover supporting domains (finance primers). The second issue was scope drift (i.e. "*Sometimes, genAI generates irrelevant information, making it challenging to identify and extract the most useful information for our needs.*"). This might be addressed through tightening prompt rules and instructing the genAI to stick to the topic and review the information for relevance with regards to the question asked. Finally, some of the genAI responses received a lower relevance rating due to boilerplate (introductions/summaries) that added text without advancing the answer. To cater to this issue genAI should omit introductions by default (P2 Constrain Verbosity, §5.3), which aligns with multimedia learning's redundancy guidance that extra, non-informative text can hinder processing [19].

**Self-efficacy.** The cohort participating in our case study reported significantly increased confidence in their ability for finance-domain upskilling and for applying DDD in their team project. Student reflections highlight preparedness and efficiency (e.g., "*understand the concept before applying it*"; "*saved me time to learn new concepts*") and that outputs triggered deeper inquiry (e.g., "*it made me ask more—and understand the context better*"). Although the reported self-efficacy gains were large, we hypothesize that low "supportiveness" of the genAI tutor—manifested as being neutral, factual and not encouraging student to explore—may limit efficacy gains for some students [20] and more pronounced results could be achieved.

## 6.2 RQ2: pedagogical value, load, support

To answer RQ2, we investigated genAI tutoring through the learning lens [21] and focused on three quality dimensions: pedagogical value, cognitive load, and supportiveness.

**Pedagogical value (depth & learning contribution).** In our sample, genAI tutoring showed high value with mean = 89.4%. Students echoed this, noting that answers were "*clear*" and helped them learn unfamiliar topics, and that "*quick, detailed answers made complex ideas easier to understand*".

The tutor was configured to generate an example, analogy, or small case study; it consistently provided ≥1 example, and for abstract topics (e.g., bounded context vs. domain service) sometimes two. For responses scoring ≥2, examples were sufficient to grasp concepts.

Two recurring issues emerged in instructor ratings and student reflections: (1) depth calibration (answers sometimes too technical or too simplified) and (2) example fit (examples occasionally mismatched the student's context). Students described "too technical" explanations that hindered quick comprehension, "too simplified" answers for complex topics, and examples that "didn't match my context."

To address these shortcomings four remedies are proposed. Firstly, provide expected granularity of information to a given query (P1 Set expected granularity, §5.3). Sample responses can be retrieved during genAI tutor prototyping and manually refined for desired granularity. Secondly, filter undesired terms to keep vocabulary aligned (P5 Block unfamiliar jargon, §5.3). In our Functional Programming course, Object Orientation terms were blocked during prototyping; we also banned advanced DDD terms (value object, aggregate) and found the instruction needed repeating in the prompt. We neglected to ban some advanced finance terms that surfaced in tutoring and increased complexity/lowered relevance. Arguably, the most structuring practice that can help increase learning contribution is ensuring only a limited amount of model examples are used (P3 Cap examples, §5.3). Research shows too many or overly generic examples can backfire as expertise grows [22, 23]. As quality of examples varied greatly in our study, pre-selecting exemplars during prototyping is crucial.

**Cognitive load (clarity, conciseness and accessibility).** With the tutor instructed to "focus on making the course content accessible," responses were generally well structured and to the point (mean = 82.78%). It is important to note there were 4 occurrences of very verbose answers (5%) and in one of those cases a student explicitly requested rephrasing. Furthermore, three reoccurring issues were identified: 1) repetition across intro/body/summary or within sections, 2) unexplained terms not in the provided knowledge base, 3) too many examples for a single prompt. Students also pointed out a need for "*less verbosity in the answers.*" Remedies (P2, P3, P5, P8 Signal, don't repeat, §5.3) collectively reduce extraneous load per Cognitive Load Theory [24].

**Support and encouragement.** Based on our findings (mean of 37.78% for supportiveness) genAI tutoring remains largely neutral and devoid of motivational language as default. Interestingly, all occurrences of encouragement took place when the prescribed prompt stated that a student accomplished something (completed event storming - prompt 3.2). A simple fix here is instructing the genAI tutor to embody a specific persona (P6 Add a supportive tone, §5.3). Low supportiveness may under-serve competence/relatedness needs and adding a light coaching tone can improve motivational climate [25] without extra length.

From Domain Understanding to Design Readiness: a playbook for GenAI-supported learning in Software EngineeringTable 3: High-leverage practices selected by the authors based on (i) observed failure modes/impact, (ii) implementation cost, and (iii) expected generalizability across SE courses.

| ID | Practice | Targets | Evidence link | Effort | Priority |
|---|---|---|---|---|---|
| P7 | Curate the knowledge base | Accuracy, Relevance | §5.2 (relevance dips), §6 (accuracy) | Medium | ★★★ |
| P2 | Constrain verbosity | Cognitive load, Relevance | §5.2 (load issues, boilerplate) | Low | ★★★ |
| P8 | Signal, don't repeat | Cognitive load | §5.2 (repetition) | Low | ★★★ |
| P9 | AI nudge | Pedagogical value, Cognitive load | §5.2 (engagement) | Low | ★★★ |
| P4 | Curate guardrail examples | Accuracy, Pedagogical value | §6 (example mismatch) | Medium | ★★★ |

### 6.3 RQ3: course and workflow levers

To maximize effectiveness at the course and workflow level, the levers that matter most are credit + LO (P12, P10), lightweight accountability (P14) and an effective genAI usage workshop (P11). Although all study participants took an interactive genAI literacy module (see 3.1.1), students expressed a need for more training, in particular how to curate effective prompts. A hands-on workshop demonstrating effective follow up paired with explicit prompting instructions (P9) could alleviate this. A small grade allocation paired with an explicit learning outcome (P12 + P10) sustains participation and legitimizes time-on-task, but it can also incentivize superficial usage. We counter this by tying points to quality rather than volume via a simple knowledge-check rubric (P15) that rewards clear reasoning, correct curation of genAI output, and at least one purposeful follow-up per student/team. Brief logs/notes (P14) provide just enough auditability and support reflection, though they introduce overhead and mild privacy concerns; both are mitigated with a one-page export template, minimal retention, and opt-out for sensitive content. TA delegation with quick calibration (P16) makes the workflow feasible beyond small cohorts but raises rater-drift risk; a 20-minute norming on a handful of anchor examples plus periodic spot checks keeps scores aligned. Other course/workflow practices (e.g., centralizing guidance; varying difficulty/format) are useful but largely self-explanatory and are omitted here for brevity.

## 7 Lessons learned & adoption path

### 7.1 What worked (highest leverage)

A course-grounded genAI tutor can provide consistently usable support for short, high-intensity design milestones when two conditions hold: (i) the tutor is instructed to leverage course-specific materials as primary sources (project specification, lecture slides, exemplar artifacts) and (ii) prompt-level guardrails reduce predictable failure modes (verbosity, jargon drift, and uncontrolled example proliferation). In our deployment, these lightweight constraints were sufficient to achieve high answer quality on core dimensions (accuracy, relevance, pedagogical value) while keeping instructor overhead low. Table 3 summarizes the highest-leverage practices we recommend for similar course settings and section 5.3 outlines operational practices that can be deployed in any genAI-based tutoring context.

### 7.2 What did not work by default and what we would change

Even with a grounded configuration, students reported that interactions sometimes required effort to "steer" the tutor: some answers were "too general" for students who already understood the topic, while others became "too technical" for quick comprehension. Students also highlighted that follow-ups were valuable but not always easy to use effectively: they appreciated immediate back-and-forth (e.g., "clarification and feedback immediately") and noted that follow-ups helped with workshop preparation (e.g., "follow-ups made event storming clearer"), yet several asked for explicit depth controls (beginner/advanced modes) and scaffolds on how to phrase follow-ups. In a repeat run, we would therefore make follow-up guidance explicit in the workflow (suggest follow-up types: visible clarify / probe / request example) and direct the tutor to build on the conversation (P9 GenAI Nudge) so the next turn is purposeful and interaction stays constructive [26]. Finally, the default tutor persona was consistently rated low on supportiveness. In a repeat run, we would explicitly prompt the model to adopt a light coaching tone that validates student attempts and offers encouragement (e.g. "*Make sure to validate student's questions and their desire to learn. This should be an opening sentence of your response*.") to meet learners' relatedness needs.

### 7.3 Adoption path

Start by introducing a small course credit (1–2%) and an explicit learning objective, then enforce a simple rubric (P15) and require one purposeful follow-up per student/team. As enrollment grows (≈30+), add a brief log template (P14) and TA calibration (P16) to preserve scoring consistency and tutoring quality at scale.

## 8  Threats to validity

**Internal validity.** Our inferences are associative, not causal. Gains may reflect co-interventions (mini-lectures, in-class activities, ad-hoc instructor/TA help) and grade incentives, not only the genAI tutor. Usage heterogeneity (how much and how well students used the tutor) and genAI proficiency likely moderated effects, and novelty may have temporarily boosted engagement. We did not measure students' prior genAI tool experience beyond completion of the university genAI literacy module; baseline proficiency may therefore confound perceived usefulness of the tool. We mitigated some threats via a standardized workflow with shared prompts and instructions, and the analyzed sample showed full adherence to this process.

**Construct validity.** Self-efficacy captures confidence rather than competence, so improvements need not equal skill gains. Response quality was scored mainly by the instructor; although TAs double-coded a subset as a consistency check, we did not collect sufficient overlapping ratings to compute reliable inter-rater reliability for the full sample. Several dimensions (e.g., accuracy, relevance) exhibited ceiling effects, reducing sensitivity to differences.

**Conclusion validity.** The item-level analysis drew a ~34.5% random sample that was evenly allocated across sessions and covered a contiguous block of prescribed prompts to preserve context; however, the sample was not stratified by prompt type, so residual sampling bias is possible.

**External validity.** Findings come from a single institution and cohort in a specific SE context (DDD + finance) using a customized ChatGPT and a curated knowledge base over a short time window. As such, generalizability is context specific.

## 9  Conclusions

In this classroom deployment, a course-grounded genAI-supported interactions produced consistently accurate and relevant responses during a fast-paced modeling milestone, with high pedagogical value and manageable cognitive load, and students reported large confidence gains for genAI-assisted domain learning and for applying DDD on a collaborative project. The main weaknesses were low supportiveness and occasional scope/verbosity issues. We translate these observations into an instructor's playbook: anchor the tutor in a curated knowledge base; set expected granularity; constrain verbosity and cap examples; pre-vet guardrail examples; and operationalize adoption with small credit tied to a simple quality rubric, brief logs, and TA calibration. Overall, genAI tutoring is a feasible, scalable complement to instruction when paired with a few low-cost guardrails.

**Future directions.** Our findings are context-bound (master's cohort, modeling+finance knowledge). Next steps are (i) replication in other SE topics and class sizes, (ii) evaluating non-scripted follow-ups for quality versus load, and (iv) automated quality & trust. Despite high observed accuracy, students reported residual distrust. We will pilot an genAI-assisted scoring pass (~20–30 answers per term) aligned to our rubric, with instructor spot-checks to gauge agreement. The aim is triage, not automation—flag potentially low-quality answers for review so students can rely on prescribed tutoring without heavy overhead.

## ACKNOWLEDGMENTS

This research was funded as part of the Generative Artificial Intelligence Teaching as Research (GAITAR) fellowship. We gratefully acknowledge the Eberly team for their invaluable support, which was instrumental in bringing this project to fruition.